\documentclass[aps, pra, twocolumn, superscriptaddress, showpacs]{revtex4-1}
\usepackage[ansinew]{inputenc}
\usepackage{epsfig}
\usepackage{graphicx}
\usepackage{color}
\usepackage{bm}
\usepackage{multirow}
\usepackage{mathrsfs}
\usepackage{revsymb}
\usepackage{amssymb}
\usepackage{amsmath}
\usepackage{slashed}
\usepackage{longtable}
\usepackage{dcolumn}
\newcolumntype{d}[1]{D{.}{.}{#1}}

\begin{document}

\title{Extraction of the electron mass from $g$~factor measurements on light hydrogenlike~ions}

\author{J. Zatorski}
\email[]{zatorski@mpi-hd.mpg.de}
\affiliation{Max Planck Institute for Nuclear Physics, Saupfercheckweg~1, 69117 Heidelberg, Germany}
\author{B. Sikora}
\affiliation{Max Planck Institute for Nuclear Physics, Saupfercheckweg~1, 69117 Heidelberg, Germany}
\author{S.~G.  Karshenboim}
\affiliation{Pulkovo Observatory, 196140 St. Petersburg, Russia}
\affiliation{Max Planck Institute for Quantum Optics, 85748 Garching, Germany}
\affiliation{Ludwig-Maximilians-Universit{\"a}t, Fakult{\"a}t f\"ur Physik, 80799 M\"unchen, Germany}
\author{S. Sturm}
\affiliation{Max Planck Institute for Nuclear Physics, Saupfercheckweg~1, 69117 Heidelberg, Germany}
\author{F. K\"ohler-Langes}
\affiliation{Max Planck Institute for Nuclear Physics, Saupfercheckweg~1, 69117 Heidelberg, Germany}
\author{K. Blaum}
\affiliation{Max Planck Institute for Nuclear Physics, Saupfercheckweg~1, 69117 Heidelberg, Germany}
\author{C. H. Keitel}
\affiliation{Max Planck Institute for Nuclear Physics, Saupfercheckweg~1, 69117 Heidelberg, Germany}
\author{Z. Harman}
\email[]{harman@mpi-hd.mpg.de}
\affiliation{Max Planck Institute for Nuclear Physics, Saupfercheckweg~1, 69117 Heidelberg, Germany}

\date{\today}

\begin{abstract}

The determination of the electron mass from Penning-trap measurements with ${}^{12}$C$^{5+}$ ions and from theoretical results for the
bound-electron $g$~factor is described in detail. Some recently calculated contributions slightly shift the extracted mass value.
Prospects of a further improvement of the electron mass are discussed both from the experimental and from the theoretical point of view.
Measurements with ${}^{4}$He$^+$ ions will enable a consistency check of the electron mass value, and in future an improvement of the $^4$He
nuclear mass and a determination of the fine-structure constant.

\end{abstract}

 \maketitle

\section{Introduction}

Recent years have seen a fast progress in the theoretical understanding and experimental precision of bound-electron $g$~factors
\cite{Sturm11,Pachucki05,Pachucki04,Verdu04,Yerokhin02,Shabaev02,Beier00,Haffner00}. It has also become possible to determine the atomic mass of the
electron $m_e$ in Penning trap $g$ factor experiments with light one-electron ions by means of the continuous Stern-Gerlach effect~\cite{Haffner00,Beier01}.
The most accurate value~\cite{Sturm14,Kohler15} of $m_e$ has been obtained from a recent measurement employing $^{12}$C$^{5+}$ ions.

For an electron bound to an ion and subjected to an external magnetic field of strength $B$, the Larmor frequency between the Zeeman sublevels
depends on the electron's magnetic moment $\mu$ by the well-known formula
\begin{equation}
\label{eq:Larmor}
\omega_{\rm L} = \frac{ 2 \mu}{\hbar} B = \frac{g}{2} \frac{e}{m_e} B\,,
\end{equation}
with $e$ being the (positive) unit charge, and $g$ the bound electron's gyromagnetic or $g$ factor. Calibrating the magnetic field
at the very position of the ion becomes possible through a measurement of the frequency of the cyclotron motion
of the ion as a whole,
\begin{equation}
\label{eq:cycl}
\omega_{\rm c} = \frac{Q}{M} B \,,
\end{equation}
where $Q$ and $M$ are the charge and mass of the one-electron ion, respectively. Combining the two above equations, the electron mass
can be expressed in units of the ion's mass as
\begin{equation}
\label{eq:massdet}
m_e = \frac{g}{2} \frac{e}{Q} \frac{\omega_{\rm c}}{\omega_{\rm L}} M\,,
\end{equation}
where we assign to the $g$~factor its theoretical value $g_{\rm theo}$. The experimentally determined quantity is the
frequency ratio $\Gamma={\omega_{\rm L}}/{\omega_{\rm c}}$. As it is clear from Eq.~(\ref{eq:massdet}), for extracting
$m_e$ to a given level of relative uncertainty, $g_{\rm theo}$, $\Gamma$ and $M$ need to be known at the same level of relative
uncertainty or better.

In our experiment~\cite{Sturm14,Kohler15} and in an earlier study~\cite{Haffner00}, $^{12}$C$^{5+}$ ions were employed since the
$^{12}$C atom defines the atomic mass unit, and, therefore, also the mass of the ion is known exceptionally well.
Our experiment has been presented in detail in Ref.~\cite{Kohler15}. In the current article, we describe theoretical details of the
extraction of the electron mass, and present a reevaluation which takes into account newly calculated quantum electrodynamic
(QED) corrections. In addition, prospects of a further improvement of the electron mass value are discussed, by employing either hydrogenlike
$^{12}$C$^{5+}$ or $^{4}$He$^{+}$ ions. A measurement with $^{4}$He$^{+}$ also enables in principle a determination of the $^{4}$He
mass, and in future the fine-structure constant~$\alpha$.

\section{Evaluation of theoretical contributions for the determination of the electron mass}

\begin{table*}[!tb]
\caption{Values of individual contributions to $g\left(\null^{4}\textrm{He}^{+}\right)$, $g\left(\null^{12}\textrm{C}^{5+}\right)$
and $g\left({}^{28}\textrm{Si}^{13+}\right)$, and some relevant nuclear parameters. The abbreviations stand for: $M_{\rm atom}$, $M$:
mass of the atom and the hydrogenlike ion, respectively; $\left\langle r^2 \right\rangle^{1/2}$: root-mean-square nuclear charge radius;
``SE'' -- self-energy correction; ``SE-FS'' -- mixed self-energy and nuclear finite size correction, ``VP-EL'' -- electric-loop
vacuum polarization correction, ``VP-ML'' -- magnetic-loop vacuum polarization correction. An experimental value for $g\left({}^{28}\textrm{Si}^{13+}\right)$
is given as published in Ref.~\cite{Sturm13}, i.e. evaluated with a former, less accurate value of $m_e$, which defines the last error given.
See text for details.}
\label{tab:table1}
\centering
\begin{ruledtabular}
\begin{tabular}{lllll}
Contribution                                          & $\null^{4}\textrm{He}^{+}$                & $\null^{12}\textrm{C}^{5+}$              & $\null^{28}\textrm{Si}^{13+}$        & Ref.                \\
\hline
$\left\langle r^2 \right\rangle^{1/2}$~[fm]           & \phantom{-}1.681(4)\footnotemark[1]       & \phantom{-}2.4703(22)                    & \phantom{-}3.1223(24)                & \cite{Angeli2013}   \\
$M_{\textrm{atom}}$~[u]                               & \phantom{-}4.002\,603\,254\,130(63)       & \phantom{-}12 (exact)                    & \phantom{-}27.976\,926\,534\,65(44)  & \cite{AME2012}   \\
$M$~[u]                                               & \phantom{-}4.002\,054\,700\,617(63)       & \phantom{-}11.997\,257\,680\,293\,69(97) & \phantom{-}27.969\,800\,594\,24(50)  & \vspace{1mm}        \\
Dirac value                                           & \phantom{-}1.999\,857\,988 825\,37(7)     & \phantom{-}1.998\,721\,354\,392\,0(6)    & \phantom{-}1.993\,023\,571\,557(3)   & \cite{Breit1928}    \\
Finite nuclear size                                   & \phantom{-}0.000\,000\,000\,002\,30(1)    & \phantom{-}0.000\,000\,000\,407\,4(7)    & \phantom{-}0.000\,000\,020\,468(31)  & \cite{Zatorski12}   \\
One-loop QED                                          &                                           &                                          &                                      &                     \\
\phantom{aaaa}   $(Z\alpha)^0$                        & \phantom{-}0.002\,322\,819\,464\,85(54)   & \phantom{-}0.002\,322\,819\,464\,9(5)    & \phantom{-}0.002\,322\,819\,465(1)   & \cite{Schwinger48,CODATA2014}   \\
\phantom{aaaa}   $(Z\alpha)^2$                        & \phantom{-}0.000\,000\,082\,462\,19       & \phantom{-}0.000\,000\,742\,159\,7       & \phantom{-}0.000\,004\,040\,647      & \cite{Grotch70}     \\
\phantom{aaaa}   $(Z\alpha)^4$                        & \phantom{-}0.000\,000\,001\,976\,70       & \phantom{-}0.000\,000\,093\,422\,2       & \phantom{-}0.000\,001\,244\,596      & \cite{Pachucki05}   \\
\phantom{aaaa}   $(Z \alpha)^{5+}$ SE                 & \phantom{-}0.000\,000\,000\,035\,42(68)   & \phantom{-}0.000\,000\,008\,282\,6(37)   & \phantom{-}0.000\,000\,542\,856(60)  & \cite{Yerokhin02,Yerokhin04}\footnotemark[2] \\
\phantom{aaaa}   SE FS                                & -0.000\,000\,000\,000\,00                 & -0.000\,000\,000\,000\,7                 & -0.000\,000\,000\,068                & \cite{YerokhinFS13} \\
\phantom{aaaa}   $\geq (Z \alpha)^5$ VP-EL            & \phantom{-}0.000\,000\,000\,002\,52       & \phantom{-}0.000\,000\,000\,555\,9       & \phantom{-}0.000\,000\,032\,531      & \cite{YerokhinFS13,Karshenboim01}      \\
\phantom{aaaa}   VP-EL FS                             & \phantom{-}0.000\,000\,000\,000\,00       & \phantom{-}0.000\,000\,000\,000\,2       & \phantom{-}0.000\,000\,000\,022      & \cite{YerokhinFS13}   \\
\phantom{aaaa}   $(Z \alpha)^{5+}$ VP-ML              & \phantom{-}0.000\,000\,000\,000\,16       & \phantom{-}0.000\,000\,000\,038\,1       & \phantom{-}0.000\,000\,002\,540(10)  & \cite{Lee05,Karshenboim02}        \\
\phantom{aaaa}   $(Z \alpha)^{5+}$ VP-ML FS           & \phantom{-}0.000\,000\,000\,000\,00       & \phantom{-}0.000\,000\,000\,000\,0       & -0.000\,000\,000\,001                & \cite{YerokhinFS13,Lee05}        \\
Two-loop QED                                          &                                           &                                          &                                      &                     \\
\phantom{aaaa}  $(Z\alpha)^0$                         & -0.000\,003\,544\,604\,49                 & -0.000\,003\,544\,604\,5                 & -0.000\,003\,544\,604                & \cite{Peterman57,Sommerfield58}   \\
\phantom{aaaa}  $(Z\alpha)^2$                         & -0.000\,000\,000\,125\,84                 & -0.000\,000\,001\,132\,5                 & -0.000\,000\,006\,166                & \cite{Grotch70}     \\
\phantom{aaaa}  $(Z\alpha)^4$ (w/o LBL)               & \phantom{-}0.000\,000\,000\,002\,41       & \phantom{-}0.000\,000\,000\,060\,1       & -0.000\,000\,001\,318                & \cite{Pachucki05,Pachucki04}   \\
\phantom{aaaa}  LBL at $(Z\alpha)^4$                  & -0.000\,000\,000\,000\,39                 & -0.000\,000\,000\,031\,5                 & -0.000\,000\,000\,933                & \cite{Czarnecki16}  \\
\phantom{aaaa} $(Z \alpha)^{5+}$ S(VP)E               & \phantom{-}0.000\,000\,000\,000\,00       & \phantom{-}0.000\,000\,000\,000\,0(1)    & \phantom{-}0.000\,000\,000\,009(2)   & \cite{Yerokhin2Loop2013} \\
\phantom{aaaa} $(Z \alpha)^{5+}$ SEVP                 & \phantom{-}0.000\,000\,000\,000\,03       & \phantom{-}0.000\,000\,000\,006\,9(3)    & \phantom{-}0.000\,000\,000\,458(1)   & \cite{Yerokhin2Loop2013} \\
\phantom{aaaa} $(Z \alpha)^{5+}$ VPVP                 & \phantom{-}0.000\,000\,000\,000\,03       & \phantom{-}0.000\,000\,000\,005\,5       & \phantom{-}0.000\,000\,000\,315      & \cite{Yerokhin2Loop2013,Jentschura09} \\
\phantom{aaaa} $(Z \alpha)^{5+}$ SESE (estimate)      & \phantom{-}0.000\,000\,000\,000\,00(2)    & -0.000\,000\,000\,001\,2(33)             & -0.000\,000\,000\,082(139)           &                     \\
$\geq$ Three-loop QED                                 &                                           &                                          &                                      &                     \\
\phantom{aaaa}  $(Z\alpha)^0$                         &  \phantom{-}0.000\,000\,029\,497\,95      & \phantom{-}0.000\,000\,029\,497\,9       & \phantom{-}0.000\,000\,029\,498      & \cite{Laporta96,Aoyama07,Aoyama12}   \\
\phantom{aaaa}  $(Z\alpha)^2$                         &  \phantom{-}0.000\,000\,000\,001\,05      & \phantom{-}0.000\,000\,000\,009\,4       & \phantom{-}0.000\,000\,000\,051      & \cite{Grotch70}     \\
Recoil                                                &                                           &                                          &                                      &                     \\
\phantom{aaaa}  $(m/M)^1$  all-orders in $(Z \alpha)$ &   \phantom{-}0.000\,000\,029\,202\,51     & \phantom{-}0.000\,000\,087\,725\,1       & \phantom{-}0.000\,000\,206\,100      & \cite{Shabaev02}    \\
\phantom{aaaa}  $(m/M)^{2+}$ at  $(Z \alpha)^2$       &   -0.000\,000\,000\,012\,01               & -0.000\,000\,000\,028\,1                 & -0.000\,000\,000\,060                & \cite{Pachucki08}   \\
\phantom{aaaa}  Radiative-recoil                      &   -0.000\,000\,000\,022\,61               & -0.000\,000\,000\,067\,9                 & -0.000\,000\,000\,159                & \cite{Grotch70,Beier00}      \\
Nuclear polarizability                                &   \phantom{-}0.000\,000\,000\,000\,00     & \phantom{-}0.000\,000\,000\,000\,0       & \phantom{-}0.000\,000\,000\,000(20)  & \cite{Nefiodov02}\footnotemark[2] \\
Nuclear susceptibility                                &   \phantom{-}0.000\,000\,000\,000\,00     & \phantom{-}0.000\,000\,000\,000\,0(1)    & \phantom{-}0.000\,000\,000\,000(3)   & \cite{Jentschura06} \\
Weak interaction at $(Z\alpha)^0$                     &   \phantom{-}0.000\,000\,000\,000\,06     & \phantom{-}0.000\,000\,000\,000\,1       & \phantom{-}0.000\,000\,000\,000      & \cite{CODATA2014,Czarnecki96}   \\
Hadronic effects at $(Z\alpha)^0$                     &   \phantom{-}0.000\,000\,000\,003\,47     & \phantom{-}0.000\,000\,000\,003\,5       & \phantom{-}0.000\,000\,000\,003      & \cite{Nomura13,Kurz14,Prades10}   \\
\hline
Total w/o SESE $(Z \alpha)^5$                         &   \phantom{-}2.002\,177\,406\,711\,68(87) & \phantom{-}2.001\,041\,590\,166\,3(39)   & \phantom{-}1.995\,348\,957\,791(71)  &                     \\
Total w/ SESE $(Z \alpha)^5$ from exp.                &   \phantom{-}2.002\,177\,406\,711\,68(87) & \phantom{-}2.001\,041\,590\,165\,2(51)   & \phantom{-}1.995\,348\,957\,708(156) &                     \\
Experiment                                            &                                           &                                          & \phantom{-}1.995\,348\,959\,10(7)$_{\rm stat}$(7)$_{\rm syst}$(80)$_{m_e}$ & \cite{Sturm13}      \\
\end{tabular}
\end{ruledtabular}
\footnotetext[1]{Ref.~\cite{Sick15}.}
\footnotetext[2]{Extrapolation of the cited results.}
\end{table*}

A great variety of physical effects contribute to the theoretical value of the $g$ factor. For a free electron, i.e. at order
$(Z\alpha)^0$, the $g$~factor can be parameterized as
\begin{equation}
g_{\rm free} = 2 \left( C^{(0)} + C^{(1)} \left( \frac{\alpha}{\pi} \right) + C^{(2)} \left( \frac{\alpha}{\pi} \right)^2 + \cdots \right)\,,
\end{equation}
with the coefficients $C^{(n)}$ representing the sum of all contributing $n$-loop QED diagrams. The leading radiative correction is determined
by the Schwinger term with $C^{(1)}={1}/{2}$. For bound electrons, the above formula has to be extended with terms accounting for the
interaction with the nuclear potential. At low atomic numbers, this interaction can be taken into account by an expansion
in $Z\alpha$. Several terms in this expansion have been calculated \cite{Grotch70,Czarnecki00,Pachucki04,Pachucki05}. Above a certain level of
accuracy, non-perturbative methods in $Z\alpha$ are also required. The leading relativistic binding term is~\cite{Breit1928}
\begin{equation}
g_{\rm Dirac}-2=\frac{4}{3}\left(\sqrt{1-(Z\alpha)^2}-1\right)\,,
\end{equation}
which needs to be extended with one- to three-loop QED binding terms as well as effects originating from the nucleus, namely, the recoil contribution
and nuclear structural effects. Further small contributions from nuclear structure may arise such as the nuclear polarizability correction. A review of the
theoretical results can be found in Refs.~\cite{CODATA2014,Pachucki05}. These contributions have been benchmarked in Ref.~\cite{Sturm11} with
hydrogenlike Si$^{13+}$, where an excellent agreement of theory and experiment was stated. In Si$^{13+}$, bound-state effects are magnified
as compared to the case of C$^{5+}$ due to power scaling in $Z\alpha$. Therefore, one can rely on the correctness of theory for C$^{5+}$ when
extracting the electron mass via Eq.~(\ref{eq:massdet}).

The experiment on $^{28}$Si$^{13+}$ was repeated later with a significantly improved precision~\cite{Sturm13}, triggering a further
advancement in the theoretical treatment. Non-perturbative (with respect to $Z\alpha$) results for a subset of two-loop QED corrections
have been published~\cite{Yerokhin2Loop2013}. In that article, the higher-order remainder in $Z\alpha$ of two-loop corrections with one
or two closed fermionic loops have been calculated in the Uehling approximation. The coefficient of the fifth-order term in $Z\alpha$
for the two-loop vacuum polarization diagrams has been evaluated in Ref.~\cite{Jentschura09}. In an even more recent
publication~\cite{Czarnecki16}, a virtual light-by-light scattering correction of order ${\alpha}^2 (Z \alpha)^4$,
which was neglected in a previous calculation~\cite{Pachucki05,Pachucki04}, has been determined. The coefficient of the term was found to be
unexpectedly large. In contrast to the evaluation of Ref.~\cite{Sturm14}, here we also take into account these new terms in the
determination of $m_e$. Table~\ref{tab:table1} lists individual theoretical contributions for hydrogenlike He$^{+}$, C$^{5+}$ and
Si$^{13+}$.

The remaining unknown two-loop self-energy correction at orders higher than $(Z \alpha)^4$, which we denote by
$g_{\textrm{2L}}^{\textrm{SE}}(Z)$, is a major challenge for theory and thus has not been evaluated yet.
One may obtain an estimation of the effect for He and C ions by means of extraction of $g_{\textrm{2L}}^{\textrm{SE}}(Z=14)$
from comparison of the theory and the experimental result for Si and subsequently rescaling it from $Z=14$ to $Z=2$ and
$Z=6$, respectively. In analogy to the corresponding Lamb shift contribution, the higher-order two-loop QED effect is assumed
to be described by the formula
\begin{align}
\label{bnlexpansion}
g_{\textrm{SESE}}(Z) =& \left( \frac{\alpha}{\pi}\right)^2 (Z \alpha)^5 \biggl\{ b_{50} + b_{63} \cdot (Z\alpha) L^3 +  \\
&b_{62} \cdot (Z\alpha) L^2 + b_{61} \cdot (Z\alpha) L + b_{60} \cdot (Z\alpha) + \ldots \biggr\}, \nonumber
\end{align}
where $L = \ln\left[(Z\alpha)^{-2}\right]$ and terms of higher order with respect to $Z\alpha$ are not taken into account. In the notation
for the $b_{nl}$ coefficients, $n$ denotes the power of $Z\alpha$ and $l$ is the power of the logarithmic term.
The expansion coefficients with $n\geq 5$ have not been calculated thus far. Formally, the leading contribution to the right-hand side of
Eq.~(\ref{bnlexpansion}) is related to $b_{50}$, but, in principle, the logarithmically enhanced terms of the next order may also be
significant.

We determine $b_{50}$ as follows: First, we restrict ourselves to the leading term in Eq.~(\ref{bnlexpansion}) which only includes the $b_{50}$
parameter. Then, a comparison of the experimental and theoretical value reads
\begin{equation}
\label{b50definition}
g_{\textrm{exp}} \left(Z\right) = g^{*}_{\textrm{th}}\left(Z\right) + \left( \frac{\alpha}{\pi}\right)^2 (Z \, \alpha)^5 \, b_{50} \, ,
\end{equation}
where $g^{*}_{\textrm{th}}(Z)$ denotes the theoretical prediction for the $g$ factor including only the known corrections, i.e.,
without $g_{\textrm{SESE}}(Z)$. The relation between $g_{\textrm{exp}} \left(Z\right)$ and the frequency ratio
$\Gamma$ determined in an experiment follows from Eq.~(\ref{eq:Larmor}) and (\ref{eq:cycl}),
\begin{equation}
g_{\textrm{exp}} \left(Z\right) = 2\frac{Q}{e} \frac{m_e}{M} \, \Gamma \,,
\end{equation}
with ${Q}/{e}=Z-1$, and employing it along with Eq.~(\ref{b50definition}) we obtain a set of equations for C and Si, namely,
\begin{align}
\label{set-Eq1}
\left( \frac{\alpha}{\pi}\right)^2 (6 \, \alpha)^5 \, b_{50} &=  \frac{10 m_e}{M(\null^{12}\textrm{C}^{5+})} \, \Gamma(\null^{12}\textrm{C}^{5+}) - g^{*}_{\textrm{th}}\left(6\right) \, ,\\
\label{set-Eq2}
\left( \frac{\alpha}{\pi}\right)^2 (14 \, \alpha)^5 \, b_{50} &= \frac{26 m_e}{M(\null^{28}\textrm{Si}^{13+})} \, \Gamma(\null^{28}\textrm{Si}^{13+}) - g^{*}_{\textrm{th}}\left(14\right) \, ,
\end{align}
with the ions' masses depending on the electron mass through the formula
\begin{eqnarray}
M\left(\null^{A}\textrm{X}^{q+}\right) &=& M\left(\null^{A}\textrm{X}\right) - (Z-1) m_e \\
 &&+ \left|E_{b}\left(\null^{A}\textrm{X}\right) - E_{b}\left(\null^{A}\textrm{X}^{q+}\right)\right| \,, \nonumber
\end{eqnarray}
where $E_{b}\left(\null^{A}\textrm{X}\right)$ is the binding energy of electrons in an atom $\textrm{X}$, expressed in unified atomic mass
units (u), and $E_{b}\left(\null^{A}\textrm{X}^{q+}\right)$ is the binding energy of the electrons in an ion $\null^{A}\textrm{X}^{q+}$, also
in u. Specifically, binding energies for $^{12}$C$^{5+}$ ions can be found in Ref.~\cite{CODATA2005}, whereas for $^{28}$Si$^{13+}$ ions
in Ref.~\cite{Martin83}. For the purpose of our calculation, it is sufficient to substitute some old value of the
electron mass (e.g. from Ref.~\cite{CODATA2012}) in the above formula since it is small compared to the nuclear mass. Therefore, we can treat the ions' masses on
the right-hand sides of Eqs.~(\ref{set-Eq1}-\ref{set-Eq2}) as known parameters. Those equations can then readily be solved for the variables
$m_e$ and $b_{50}$, namely,
\begin{widetext}
\begin{align}
\label{set2-Eq1}
m_e &= \frac{\left[ 243 \, g^{*}_{\textrm{th}}\left(14\right) - 16807 g^{*}_{\textrm{th}}\left(6\right) \right]
M(\null^{12}\textrm{C}^{5+}) M(\null^{28}\textrm{Si}^{13+})}{2 \left[ 3159 M(\null^{12}\textrm{C}^{5+})
\Gamma(\null^{28}\textrm{Si}^{13+}) - 84035 M(\null^{28}\textrm{Si}^{13+})  \Gamma(\null^{12}\textrm{C}^{5+}) \right]}  \, ,\\
\label{set2-Eq2}
 b_{50} &= \frac{\pi^2 \left[ 13 g^{*}_{\textrm{th}}\left(6\right) M(\null^{12}\textrm{C}^{5+}) \Gamma(\null^{28}\textrm{Si}^{13+}) -
 5 g^{*}_{\textrm{th}}\left(14\right) M(\null^{28}\textrm{Si}^{13+}) \Gamma(\null^{12}\textrm{C}^{5+}) \right]}
 {32 \alpha^7 \left[ 84035 M(\null^{28}\textrm{Si}^{13+})  \Gamma(\null^{12}\textrm{C}^{5+}) - 3159 M(\null^{12}\textrm{C}^{5+})
\Gamma(\null^{28}\textrm{Si}^{13+})\right]}  \, .
\end{align}
\end{widetext}
An obvious source of uncertainty of our value of $m_e$ originates from the uncertainties of the quantities in Eqs.~(\ref{set2-Eq1})
and~(\ref{set2-Eq2}). That contribution can be obtained according to the standard error propagation formula. The theory values occurring
in Eqs.~(\ref{set2-Eq1}),~(\ref{set2-Eq2}) together with their uncertainties read:
$g^{*}_{\textrm{th}}\left(\null^{12}\textrm{C}^{5+}\right) = 2.001\,041\,590\,166\,3(39)$,
$g^{*}_{\textrm{th}}\left(\null^{28}\textrm{Si}^{13+}\right) = 1.995\,348\,957\,791(71)$.
The uncertainties of relevant contributing corrections can be found in Table \ref{tab:table1}. The ion masses are given in
Table~\ref{tab:table1}, the inverse of the fine-structure constant occurring in Eqs.~(\ref{set2-Eq1}),~(\ref{set2-Eq2}) is
$\alpha^{-1} = 137.035\,999\,139(31)$~\cite{CODATA2014}, and the experimental values are
$\Gamma(\null^{12}\textrm{C}^{5+}) = 4376.210\,500\,872 (102)(69)$ (Ref.~\cite{Kohler15}) and
$\Gamma(\null^{28}\textrm{Si}^{13+}) = 3912.866\,064\,99(13)(13)$ (Ref.~\cite{Sturm13}).
The absolute electron mass uncertainty resulting through error propagation via Eq.~(\ref{set2-Eq1}) equals
$\delta_{\rm st} m_e = 1.57 \cdot 10^{-14}$~u.

Another source of uncertainty is the presence of unknown $b_{6k}$ parameters in Eq.~(\ref{bnlexpansion}). Clearly, one cannot rigorously fit
more than one $b$ parameter since one has only two equations at hand. Therefore, we tested various configurations of the $b'$s to asses the
sensitivity of our results due to changes of these parameters. Our estimation obtained this way is $\delta_{b} m_e = 8 \cdot 10^{-16}$~u.
This uncertainty was linearly added to $\delta_{\textrm{st}} m_e$.
Our final value for the electron mass reads
\begin{equation}
m_e = 0.000\,548\,579\,909\,065(16)~\mbox{u}.
\end{equation}
This value is shifted upward by 0.3~$\sigma$ with respect to earlier evaluations of the same experimental data~\cite{Kohler15,CODATA2014}
due to the inclusion of light-by-light scattering terms of order ${\alpha}^2 (Z \alpha)^4$~\cite{Czarnecki16}.

\section{Further possible improvements}

Currently, the relative uncertainty of $g_{\rm theo}$ for C$^{5+}$ is an order of magnitude better than that of $\Gamma$, i.e. it does not
hinder an improvement of $m_e$. A further enhancement of the accuracy of the experimental frequency ratio $\Gamma$ is expected for any ion
from the currently commissioned Penning-trap setup ALPHATRAP at the Max Planck Institute for Nuclear Physics~\cite{Sturm17,Sturm13AP}.
Presently, the main limitations for such measurements are the interaction of the ion with the trap electrodes ("image charge shift")
and the thermal distribution of the ion's kinetic energy. The ALPHATRAP setup will drastically reduce both effects. A larger trap diameter
decreases the image charge effect by almost two orders of magnitude compared to the Mainz $g$ factor experiment, and sympathetic laser cooling
of the highly charged ions can potentially eliminate the limitation arising from the thermal distribution. Combined, these improvements pave
the way for a significant  -- approximately one order of magnitude -- improvement in the measurement of the $g$ factor especially of light ions.

Table~\ref{tab:table1} shows that on the theoretical side, the main limitation arises from the accuracy of one-loop SE terms of order
$(Z\alpha)^5$ and higher, which have been extracted from numerical calculations~\cite{Yerokhin02,Yerokhin04}. For low charge numbers, such calculations
are restricted by severe numerical cancellations. A significant improvement will nevertheless be possible in the nearest future~\cite{Yerokhin17}.

Another possibility to determine $m_e$ could be to employ an even lighter hydrogenlike ion, where QED binding corrections are further scaled down.
At the current level of experimental accuracy, the lightest such ion, namely, He$^+$ would deliver a valuable consistency check of the electron mass
determination. One may extract $m_e$ from a He$^+$ measurement just as accurately as from C$^{5+}$, assuming the same fractional accuracy of $\Gamma$
in both experiments. A combined analysis including the He$^+$ data and an accordingly extended system of equations [see Eq.~(\ref{set-Eq1}) and
(\ref{set-Eq2})] would lead to a slightly reduced $m_e$ uncertainty even at the present level of experimental accuracy.
The theoretical value of the $g$ factor has a significantly better relative accuracy for He$^+$ than for C$^{5+}$ due to power scaling: e.g., terms
of order $(Z\alpha)^5$ are scaled down by a factor of $3^5=243$, i.e. by more than two orders of magnitude. In case of He$^+$, therefore,
there is no need to estimate the so far uncalculated higher-order two-loop terms from the Si experiment, nor include the very recently calculated
virtual light-by-light scattering contributions.

With a further improvement of experimental accuracy by, e.g., the ALPHATRAP experiment, one can further improve $m_e$
both from C$^{5+}$ as well as from He$^{+}$. However, this improvement is limited approximately to a factor of 2 with He$^+$ due
to the current relative accuracy~\cite{CODATA2014} $\delta M/M=1.6 \cdot 10^{-11}$ of the He$^+$ ion mass. The QED theory is not a
limitation yet at this level, nor the C$^{5+}$ ion mass. At an even higher level of experimental accuracy (more than a factor of 3 better
than now), a similar experiment on $^{4}$He$^+$ will allow an improved determination of the $^{4}$He$^+$ ion's mass by solving Eq.~(\ref{eq:massdet})
for $M$, provided that a corresponding improvement of the electron mass will have been achieved with C$^{5+}$ or by some alternative means.
We note that the $^{4}$He mass is also planned to be measured by the THe-Trap experiment with an anticipated fractional accuracy of
$10^{-11}$~\cite{Streubel14}.

\begin{figure}[tb]
\centering
\includegraphics[width=0.5\textwidth]{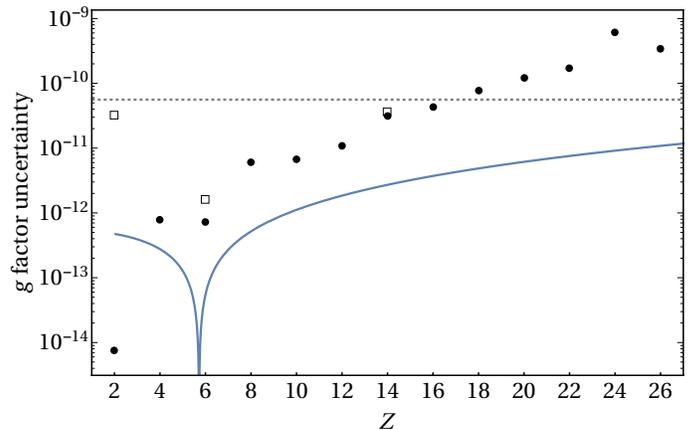}
\caption{Comparison of the uncertainties of the $g$~factor due to the uncertainty of the finite nuclear size effect
(black dots) and that of the current fine-structure constant (continuous line). $Z$ is the atomic number, and
nuclear rms charge radii and their uncertainties were taken from Ref.~\cite{Angeli2013}. The dashed line shows the current
absolute experimental error~\cite{Sturm14,Kohler15}, which at the same time determines the current error due to $m_e$.
The uncertainties due to errors of the ion masses are shown for the elements discussed (empty squares).}
\label{fig:alphasens}
\end{figure}

Let us discuss now the prospects of determining the fine-structure constant from a measurement with He$^+$. It can be extracted
from the $g$ factor, and the latter is determined by solving Eq.~(\ref{eq:massdet}) for $g$. Therefore, a competitive determination
of $\alpha$ is limited by the fractional accuracy of $m_e$, $M$, $\Gamma$, and $g_{\rm theo}$. Typically, the theoretical value of
the $g$~factor is limited by the insufficient knowledge on nuclear parameters such as the charge radius. In the light He$^+$ ion,
nuclear size effects are naturally very small; furthermore, the leading dependence on $\alpha$ does not stem from binding effects,
i.e. those scaling with some power of $Z\alpha$, but from the leading free-electron QED contribution, the Schwinger term
${\alpha}/{\pi}$. Fig.~\ref{fig:alphasens} shows that He$^+$ is the only H-like ion where the error of the $g$ factor due to the
present uncertainty of $\alpha$ is larger than the error due to the nuclear size effect. Therefore, an improved determination of
$\alpha$ is possible at all. This will require, however, an improvement in the measurement of the frequency ratio $\Gamma$ by two
orders of magnitude or better, a similar enhancement of $m_e$ from a C$^{5+}$ ion measurement or from some other source,
and an improvement of the $^4$He nuclear mass by some independent means.

We note that for other elements where the nuclear uncertainties limit the determination of the fine-structure constant, schemes have been put
forward to suppress nuclear structural effects. These contributions can be largely cancelled by appropriately chosen weighted differences
of the $g$~factor of a hydrogenlike ion and the $g$~factor corresponding to some another charge state of the same
element~\cite{Shabaev06,Yerokhin16,Yerokhin16PRA}, enabling a competitive extraction of~$\alpha$.

The determination of $\alpha$ from the $g$~factor of the lightly bound electron in He${}^{+}$ is closely related to the determination from the $g-2$
of the free electron, with the difference that theoretical binding corrections need to be subtracted first from the measured $g$~factor.
After a foreseeable improvement of the numerical accuracy of the one-loop binding self-energy correction~\cite{Yerokhin17}, such an extraction of
$\alpha$ is, from a theoretical point of view, is dominantly limited by the accuracy of free-electron QED (see Table~\ref{tab:table1}). On the
experimental side, the measurement of the bound electrons $g$ factor differs significantly from that of the free electron. In the latter case,
about three orders of magnitude in precision is gained by directly measuring $g-2\approx 0.002$ rather than $g\approx2$, which exploits the
similarity of the electrons cyclotron and Larmor frequencies. For the bound electron, these two frequencies are however very dissimilar, thus
the cyclotron frequency of the heavy ion has to be measured about three orders of magnitude more precisely to achieve a comparable precision.
An advantage employing ions may be however the large reduction of relativistic shifts, which pose a severe limitation for free electrons.

\section{Summary}

We presented an evaluation of the electron mass from Penning-trap measurements of the Larmor and cyclotron frequency ratio $\Gamma$ of a
hydrogenlike $^{12}$C$^{5+}$ ion, and the corresponding theoretical value of the bound-electron $g$~factor. So far uncalculated two-loop
self-energy corrections of order $(Z\alpha)^5$ or higher were estimated from the measured $g$ factor value of the ${}^{28}$Si$^{13+}$ ion.
This evaluation includes, in contrast to Ref.~\cite{Sturm14}, results of a non-perturbative calculation for the VPVP and SEVP
corrections~\cite{Yerokhin2Loop2013}, and a two-loop virtual light-by-light scattering contribution of order
${\alpha}^2 (Z\alpha)^4$~\cite{Czarnecki16}. The latter causes a shift of the extracted electron mass by 0.3~$\sigma$.

Prospects of further improving $m_e$ with ${}^{12}$C$^{5+}$ or $^{4}$He$^{+}$ ions were discussed. Measurements on the latter system also
allow in principle an enhanced determination of the $^{4}$He mass. A competitive determination of the fine-structure constant might be achieved
in future from a measurement with $^{4}$He$^{+}$ ions, once an experimental improvement of $\Gamma$ by two orders of magnitude becomes possible.

\section{Acknowledgments}

We acknowledge insightful conversations with A. Czarnecki and V. A. Yerokhin. This work is part of and supported by the
German Reseach Foundation (DFG) Collaborative Research Centre "SFB 1225 (ISOQUANT)". S. G. K. acknowledges support from
the DFG, grant No. KA 4645/1-1.


\begin{thebibliography}{51}
\expandafter\ifx\csname natexlab\endcsname\relax\def\natexlab#1{#1}\fi
\expandafter\ifx\csname bibnamefont\endcsname\relax
  \def\bibnamefont#1{#1}\fi
\expandafter\ifx\csname bibfnamefont\endcsname\relax
  \def\bibfnamefont#1{#1}\fi
\expandafter\ifx\csname citenamefont\endcsname\relax
  \def\citenamefont#1{#1}\fi
\expandafter\ifx\csname url\endcsname\relax
  \def\url#1{\texttt{#1}}\fi
\expandafter\ifx\csname urlprefix\endcsname\relax\def\urlprefix{URL }\fi
\providecommand{\bibinfo}[2]{#2}
\providecommand{\eprint}[2][]{\url{#2}}

\bibitem[{\citenamefont{Sturm et~al.}(2011)\citenamefont{Sturm, Wagner,
  Schabinger, Zatorski, Harman, Quint, Werth, Keitel, and Blaum}}]{Sturm11}
\bibinfo{author}{\bibfnamefont{S.}~\bibnamefont{Sturm}},
  \bibinfo{author}{\bibfnamefont{A.}~\bibnamefont{Wagner}},
  \bibinfo{author}{\bibfnamefont{B.}~\bibnamefont{Schabinger}},
  \bibinfo{author}{\bibfnamefont{J.}~\bibnamefont{Zatorski}},
  \bibinfo{author}{\bibfnamefont{Z.}~\bibnamefont{Harman}},
  \bibinfo{author}{\bibfnamefont{W.}~\bibnamefont{Quint}},
  \bibinfo{author}{\bibfnamefont{G.}~\bibnamefont{Werth}},
  \bibinfo{author}{\bibfnamefont{C.~H.} \bibnamefont{Keitel}},
  \bibnamefont{and} \bibinfo{author}{\bibfnamefont{K.}~\bibnamefont{Blaum}},
  \bibinfo{journal}{Phys. Rev. Lett.} \textbf{\bibinfo{volume}{107}},
  \bibinfo{pages}{023002} (\bibinfo{year}{2011}).

\bibitem[{\citenamefont{Pachucki et~al.}(2005)\citenamefont{Pachucki,
  Czarnecki, Jentschura, and Yerokhin}}]{Pachucki05}
\bibinfo{author}{\bibfnamefont{K.}~\bibnamefont{Pachucki}},
  \bibinfo{author}{\bibfnamefont{A.}~\bibnamefont{Czarnecki}},
  \bibinfo{author}{\bibfnamefont{U.~D.} \bibnamefont{Jentschura}},
  \bibnamefont{and} \bibinfo{author}{\bibfnamefont{V.~A.}
  \bibnamefont{Yerokhin}}, \bibinfo{journal}{Phys. Rev. A}
  \textbf{\bibinfo{volume}{72}}, \bibinfo{pages}{022108}
  (\bibinfo{year}{2005}).

\bibitem[{\citenamefont{Pachucki et~al.}(2004)\citenamefont{Pachucki,
  Jentschura, and Yerokhin}}]{Pachucki04}
\bibinfo{author}{\bibfnamefont{K.}~\bibnamefont{Pachucki}},
  \bibinfo{author}{\bibfnamefont{U.~D.} \bibnamefont{Jentschura}},
  \bibnamefont{and} \bibinfo{author}{\bibfnamefont{V.~A.}
  \bibnamefont{Yerokhin}}, \bibinfo{journal}{Phys. Rev. Lett.}
  \textbf{\bibinfo{volume}{93}}, \bibinfo{pages}{150401}
  (\bibinfo{year}{2004}).

\bibitem[{\citenamefont{Verd\'u et~al.}(2004)\citenamefont{Verd\'u,
  Djeki\ifmmode~\acute{c}\else \'{c}\fi{}, Stahl, Valenzuela, Vogel, Werth,
  Beier, Kluge, and Quint}}]{Verdu04}
\bibinfo{author}{\bibfnamefont{J.}~\bibnamefont{Verd\'u}},
  \bibinfo{author}{\bibfnamefont{S.}~\bibnamefont{Djeki\ifmmode~\acute{c}\else
  \'{c}\fi{}}}, \bibinfo{author}{\bibfnamefont{S.}~\bibnamefont{Stahl}},
  \bibinfo{author}{\bibfnamefont{T.}~\bibnamefont{Valenzuela}},
  \bibinfo{author}{\bibfnamefont{M.}~\bibnamefont{Vogel}},
  \bibinfo{author}{\bibfnamefont{G.}~\bibnamefont{Werth}},
  \bibinfo{author}{\bibfnamefont{T.}~\bibnamefont{Beier}},
  \bibinfo{author}{\bibfnamefont{H.-J.} \bibnamefont{Kluge}}, \bibnamefont{and}
  \bibinfo{author}{\bibfnamefont{W.}~\bibnamefont{Quint}},
  \bibinfo{journal}{Phys. Rev. Lett.} \textbf{\bibinfo{volume}{92}},
  \bibinfo{pages}{093002} (\bibinfo{year}{2004}).

\bibitem[{\citenamefont{Yerokhin et~al.}(2002)\citenamefont{Yerokhin,
  Indelicato, and Shabaev}}]{Yerokhin02}
\bibinfo{author}{\bibfnamefont{V.~A.} \bibnamefont{Yerokhin}},
  \bibinfo{author}{\bibfnamefont{P.}~\bibnamefont{Indelicato}},
  \bibnamefont{and} \bibinfo{author}{\bibfnamefont{V.~M.}
  \bibnamefont{Shabaev}}, \bibinfo{journal}{Phys. Rev. Lett.}
  \textbf{\bibinfo{volume}{89}}, \bibinfo{pages}{143001}
  (\bibinfo{year}{2002}).

\bibitem[{\citenamefont{Shabaev and Yerokhin}(2002)}]{Shabaev02}
\bibinfo{author}{\bibfnamefont{V.~M.} \bibnamefont{Shabaev}} \bibnamefont{and}
  \bibinfo{author}{\bibfnamefont{V.~A.} \bibnamefont{Yerokhin}},
  \bibinfo{journal}{Phys. Rev. Lett.} \textbf{\bibinfo{volume}{88}},
  \bibinfo{pages}{091801} (\bibinfo{year}{2002}).

\bibitem[{\citenamefont{Beier}(2000)}]{Beier00}
\bibinfo{author}{\bibfnamefont{T.}~\bibnamefont{Beier}},
  \bibinfo{journal}{Phys. Rep.} \textbf{\bibinfo{volume}{339}},
  \bibinfo{pages}{79 } (\bibinfo{year}{2000}).

\bibitem[{\citenamefont{H\"affner et~al.}(2000)\citenamefont{H\"affner, Beier,
  Hermanspahn, Kluge, Quint, Stahl, Verd\'u, and Werth}}]{Haffner00}
\bibinfo{author}{\bibfnamefont{H.}~\bibnamefont{H\"affner}},
  \bibinfo{author}{\bibfnamefont{T.}~\bibnamefont{Beier}},
  \bibinfo{author}{\bibfnamefont{N.}~\bibnamefont{Hermanspahn}},
  \bibinfo{author}{\bibfnamefont{H.-J.} \bibnamefont{Kluge}},
  \bibinfo{author}{\bibfnamefont{W.}~\bibnamefont{Quint}},
  \bibinfo{author}{\bibfnamefont{S.}~\bibnamefont{Stahl}},
  \bibinfo{author}{\bibfnamefont{J.}~\bibnamefont{Verd\'u}}, \bibnamefont{and}
  \bibinfo{author}{\bibfnamefont{G.}~\bibnamefont{Werth}},
  \bibinfo{journal}{Phys. Rev. Lett.} \textbf{\bibinfo{volume}{85}},
  \bibinfo{pages}{5308} (\bibinfo{year}{2000}).

\bibitem[{\citenamefont{Beier et~al.}(2001)\citenamefont{Beier, H\"affner,
  Hermanspahn, Karshenboim, Kluge, Quint, Stahl, Verd\'u, and Werth}}]{Beier01}
\bibinfo{author}{\bibfnamefont{T.}~\bibnamefont{Beier}},
  \bibinfo{author}{\bibfnamefont{H.}~\bibnamefont{H\"affner}},
  \bibinfo{author}{\bibfnamefont{N.}~\bibnamefont{Hermanspahn}},
  \bibinfo{author}{\bibfnamefont{S.~G.} \bibnamefont{Karshenboim}},
  \bibinfo{author}{\bibfnamefont{H.-J.} \bibnamefont{Kluge}},
  \bibinfo{author}{\bibfnamefont{W.}~\bibnamefont{Quint}},
  \bibinfo{author}{\bibfnamefont{S.}~\bibnamefont{Stahl}},
  \bibinfo{author}{\bibfnamefont{J.}~\bibnamefont{Verd\'u}}, \bibnamefont{and}
  \bibinfo{author}{\bibfnamefont{G.}~\bibnamefont{Werth}},
  \bibinfo{journal}{Phys. Rev. Lett.} \textbf{\bibinfo{volume}{88}},
  \bibinfo{pages}{011603} (\bibinfo{year}{2001}).

\bibitem[{\citenamefont{Sturm et~al.}(2014)\citenamefont{Sturm, K{\"o}hler,
  Zatorski, Wagner, Harman, Werth, Quint, Keitel, and Blaum}}]{Sturm14}
\bibinfo{author}{\bibfnamefont{S.}~\bibnamefont{Sturm}},
  \bibinfo{author}{\bibfnamefont{F.}~\bibnamefont{K{\"o}hler}},
  \bibinfo{author}{\bibfnamefont{J.}~\bibnamefont{Zatorski}},
  \bibinfo{author}{\bibfnamefont{A.}~\bibnamefont{Wagner}},
  \bibinfo{author}{\bibfnamefont{Z.}~\bibnamefont{Harman}},
  \bibinfo{author}{\bibfnamefont{G.}~\bibnamefont{Werth}},
  \bibinfo{author}{\bibfnamefont{W.}~\bibnamefont{Quint}},
  \bibinfo{author}{\bibfnamefont{C.~H.} \bibnamefont{Keitel}},
  \bibnamefont{and} \bibinfo{author}{\bibfnamefont{K.}~\bibnamefont{Blaum}},
  \bibinfo{journal}{Nature} \textbf{\bibinfo{volume}{506}},
  \bibinfo{pages}{467} (\bibinfo{year}{2014}).

\bibitem[{\citenamefont{K\"ohler et~al.}(2015)\citenamefont{K\"ohler, Sturm,
  Kracke, Werth, Quint, and Blaum}}]{Kohler15}
\bibinfo{author}{\bibfnamefont{F.}~\bibnamefont{K\"ohler}},
  \bibinfo{author}{\bibfnamefont{S.}~\bibnamefont{Sturm}},
  \bibinfo{author}{\bibfnamefont{A.}~\bibnamefont{Kracke}},
  \bibinfo{author}{\bibfnamefont{G.}~\bibnamefont{Werth}},
  \bibinfo{author}{\bibfnamefont{W.}~\bibnamefont{Quint}}, \bibnamefont{and}
  \bibinfo{author}{\bibfnamefont{K.}~\bibnamefont{Blaum}},
  \bibinfo{journal}{Journal of Physics B: Atomic, Molecular and Optical
  Physics} \textbf{\bibinfo{volume}{48}}, \bibinfo{pages}{144032}
  (\bibinfo{year}{2015}).

\bibitem[{\citenamefont{Sturm et~al.}(2013{\natexlab{a}})\citenamefont{Sturm,
  Wagner, Kretzschmar, Quint, Werth, and Blaum}}]{Sturm13}
\bibinfo{author}{\bibfnamefont{S.}~\bibnamefont{Sturm}},
  \bibinfo{author}{\bibfnamefont{A.}~\bibnamefont{Wagner}},
  \bibinfo{author}{\bibfnamefont{M.}~\bibnamefont{Kretzschmar}},
  \bibinfo{author}{\bibfnamefont{W.}~\bibnamefont{Quint}},
  \bibinfo{author}{\bibfnamefont{G.}~\bibnamefont{Werth}}, \bibnamefont{and}
  \bibinfo{author}{\bibfnamefont{K.}~\bibnamefont{Blaum}},
  \bibinfo{journal}{Phys. Rev. A} \textbf{\bibinfo{volume}{87}},
  \bibinfo{pages}{030501} (\bibinfo{year}{2013}{\natexlab{a}}).

\bibitem[{\citenamefont{Angeli and Marinova}(2013)}]{Angeli2013}
\bibinfo{author}{\bibfnamefont{I.}~\bibnamefont{Angeli}} \bibnamefont{and}
  \bibinfo{author}{\bibfnamefont{K.}~\bibnamefont{Marinova}},
  \bibinfo{journal}{Atomic Data and Nuclear Data Tables}
  \textbf{\bibinfo{volume}{99}}, \bibinfo{pages}{69 } (\bibinfo{year}{2013}).

\bibitem[{\citenamefont{Wang et~al.}(2012)\citenamefont{Wang, Audi, Wapstra,
  Kondev, MacCormick, Xu, and Pfeiffer}}]{AME2012}
\bibinfo{author}{\bibfnamefont{M.}~\bibnamefont{Wang}},
  \bibinfo{author}{\bibfnamefont{G.}~\bibnamefont{Audi}},
  \bibinfo{author}{\bibfnamefont{A.}~\bibnamefont{Wapstra}},
  \bibinfo{author}{\bibfnamefont{F.}~\bibnamefont{Kondev}},
  \bibinfo{author}{\bibfnamefont{M.}~\bibnamefont{MacCormick}},
  \bibinfo{author}{\bibfnamefont{X.}~\bibnamefont{Xu}}, \bibnamefont{and}
  \bibinfo{author}{\bibfnamefont{B.}~\bibnamefont{Pfeiffer}},
  \bibinfo{journal}{Chinese Physics C} \textbf{\bibinfo{volume}{36}},
  \bibinfo{pages}{1603} (\bibinfo{year}{2012}).

\bibitem[{\citenamefont{Breit}(1928)}]{Breit1928}
\bibinfo{author}{\bibfnamefont{G.}~\bibnamefont{Breit}},
  \bibinfo{journal}{Nature} \textbf{\bibinfo{volume}{122}},
  \bibinfo{pages}{649} (\bibinfo{year}{1928}).

\bibitem[{\citenamefont{Zatorski et~al.}(2012)\citenamefont{Zatorski,
  Oreshkina, Keitel, and Harman}}]{Zatorski12}
\bibinfo{author}{\bibfnamefont{J.}~\bibnamefont{Zatorski}},
  \bibinfo{author}{\bibfnamefont{N.~S.} \bibnamefont{Oreshkina}},
  \bibinfo{author}{\bibfnamefont{C.~H.} \bibnamefont{Keitel}},
  \bibnamefont{and} \bibinfo{author}{\bibfnamefont{Z.}~\bibnamefont{Harman}},
  \bibinfo{journal}{Phys. Rev. Lett.} \textbf{\bibinfo{volume}{108}},
  \bibinfo{pages}{063005} (\bibinfo{year}{2012}).

\bibitem[{\citenamefont{Schwinger}(1948)}]{Schwinger48}
\bibinfo{author}{\bibfnamefont{J.}~\bibnamefont{Schwinger}},
  \bibinfo{journal}{Phys. Rev.} \textbf{\bibinfo{volume}{73}},
  \bibinfo{pages}{416} (\bibinfo{year}{1948}).

\bibitem[{\citenamefont{Mohr et~al.}(2016)\citenamefont{Mohr, Newell, and
  Taylor}}]{CODATA2014}
\bibinfo{author}{\bibfnamefont{P.~J.} \bibnamefont{Mohr}},
  \bibinfo{author}{\bibfnamefont{D.~B.} \bibnamefont{Newell}},
  \bibnamefont{and} \bibinfo{author}{\bibfnamefont{B.~N.}
  \bibnamefont{Taylor}}, \bibinfo{journal}{Rev. Mod. Phys.}
  \textbf{\bibinfo{volume}{88}}, \bibinfo{pages}{035009}
  (\bibinfo{year}{2016}).

\bibitem[{\citenamefont{Grotch}(1970)}]{Grotch70}
\bibinfo{author}{\bibfnamefont{H.}~\bibnamefont{Grotch}},
  \bibinfo{journal}{Phys. Rev. Lett.} \textbf{\bibinfo{volume}{24}},
  \bibinfo{pages}{39} (\bibinfo{year}{1970}).

\bibitem[{\citenamefont{Yerokhin et~al.}(2004)\citenamefont{Yerokhin,
  Indelicato, and Shabaev}}]{Yerokhin04}
\bibinfo{author}{\bibfnamefont{V.~A.} \bibnamefont{Yerokhin}},
  \bibinfo{author}{\bibfnamefont{P.}~\bibnamefont{Indelicato}},
  \bibnamefont{and} \bibinfo{author}{\bibfnamefont{V.~M.}
  \bibnamefont{Shabaev}}, \bibinfo{journal}{Phys. Rev. A}
  \textbf{\bibinfo{volume}{69}}, \bibinfo{pages}{052503}
  (\bibinfo{year}{2004}).

\bibitem[{\citenamefont{Yerokhin et~al.}(2013)\citenamefont{Yerokhin, Keitel,
  and Harman}}]{YerokhinFS13}
\bibinfo{author}{\bibfnamefont{V.~A.} \bibnamefont{Yerokhin}},
  \bibinfo{author}{\bibfnamefont{C.~H.} \bibnamefont{Keitel}},
  \bibnamefont{and} \bibinfo{author}{\bibfnamefont{Z.}~\bibnamefont{Harman}},
  \bibinfo{journal}{Journal of Physics B: Atomic, Molecular and Optical
  Physics} \textbf{\bibinfo{volume}{46}}, \bibinfo{pages}{245002}
  (\bibinfo{year}{2013}).

\bibitem[{\citenamefont{Karshenboim et~al.}(2001)\citenamefont{Karshenboim,
  Ivanov, and Shabaev}}]{Karshenboim01}
\bibinfo{author}{\bibfnamefont{S.~G.} \bibnamefont{Karshenboim}},
  \bibinfo{author}{\bibfnamefont{V.~G.} \bibnamefont{Ivanov}},
  \bibnamefont{and} \bibinfo{author}{\bibfnamefont{V.~M.}
  \bibnamefont{Shabaev}}, \bibinfo{journal}{Journal of Experimental and
  Theoretical Physics} \textbf{\bibinfo{volume}{93}}, \bibinfo{pages}{477}
  (\bibinfo{year}{2001}), ISSN \bibinfo{issn}{1090-6509}.

\bibitem[{\citenamefont{Lee et~al.}(2005)\citenamefont{Lee, Milstein, Terekhov,
  and Karshenboim}}]{Lee05}
\bibinfo{author}{\bibfnamefont{R.~N.} \bibnamefont{Lee}},
  \bibinfo{author}{\bibfnamefont{A.~I.} \bibnamefont{Milstein}},
  \bibinfo{author}{\bibfnamefont{I.~S.} \bibnamefont{Terekhov}},
  \bibnamefont{and} \bibinfo{author}{\bibfnamefont{S.~G.}
  \bibnamefont{Karshenboim}}, \bibinfo{journal}{Phys. Rev. A}
  \textbf{\bibinfo{volume}{71}}, \bibinfo{pages}{052501}
  (\bibinfo{year}{2005}).

\bibitem[{\citenamefont{Karshenboim and Milstein}(2002)}]{Karshenboim02}
\bibinfo{author}{\bibfnamefont{S.~G.} \bibnamefont{Karshenboim}}
  \bibnamefont{and} \bibinfo{author}{\bibfnamefont{A.~I.}
  \bibnamefont{Milstein}}, \bibinfo{journal}{Phys. Lett. B}
  \textbf{\bibinfo{volume}{549}}, \bibinfo{pages}{321 } (\bibinfo{year}{2002}).

\bibitem[{\citenamefont{Peterman}(1957)}]{Peterman57}
\bibinfo{author}{\bibfnamefont{A.}~\bibnamefont{Peterman}},
  \bibinfo{journal}{Helv. Phys. Act} \textbf{\bibinfo{volume}{30}},
  \bibinfo{pages}{407} (\bibinfo{year}{1957}).

\bibitem[{\citenamefont{Sommerfield}(1958)}]{Sommerfield58}
\bibinfo{author}{\bibfnamefont{C.~M.} \bibnamefont{Sommerfield}},
  \bibinfo{journal}{Ann. Phys.} \textbf{\bibinfo{volume}{5}},
  \bibinfo{pages}{26 } (\bibinfo{year}{1958}).

\bibitem[{\citenamefont{Czarnecki and Szafron}(2016)}]{Czarnecki16}
\bibinfo{author}{\bibfnamefont{A.}~\bibnamefont{Czarnecki}} \bibnamefont{and}
  \bibinfo{author}{\bibfnamefont{R.}~\bibnamefont{Szafron}},
  \bibinfo{journal}{Phys. Rev. A} \textbf{\bibinfo{volume}{94}},
  \bibinfo{pages}{060501(R)} (\bibinfo{year}{2016}).

\bibitem[{\citenamefont{Yerokhin and Harman}(2013)}]{Yerokhin2Loop2013}
\bibinfo{author}{\bibfnamefont{V.~A.} \bibnamefont{Yerokhin}} \bibnamefont{and}
  \bibinfo{author}{\bibfnamefont{Z.}~\bibnamefont{Harman}},
  \bibinfo{journal}{Phys. Rev. A} \textbf{\bibinfo{volume}{88}},
  \bibinfo{pages}{042502} (\bibinfo{year}{2013}).

\bibitem[{\citenamefont{Jentschura}(2009)}]{Jentschura09}
\bibinfo{author}{\bibfnamefont{U.~D.} \bibnamefont{Jentschura}},
  \bibinfo{journal}{Phys. Rev. A} \textbf{\bibinfo{volume}{79}},
  \bibinfo{pages}{044501} (\bibinfo{year}{2009}).

\bibitem[{\citenamefont{Laporta and Remiddi}(1996)}]{Laporta96}
\bibinfo{author}{\bibfnamefont{S.}~\bibnamefont{Laporta}} \bibnamefont{and}
  \bibinfo{author}{\bibfnamefont{E.}~\bibnamefont{Remiddi}},
  \bibinfo{journal}{Physics Letters B} \textbf{\bibinfo{volume}{379}},
  \bibinfo{pages}{283 } (\bibinfo{year}{1996}).

\bibitem[{\citenamefont{Aoyama et~al.}(2007)\citenamefont{Aoyama, Hayakawa,
  Kinoshita, and Nio}}]{Aoyama07}
\bibinfo{author}{\bibfnamefont{T.}~\bibnamefont{Aoyama}},
  \bibinfo{author}{\bibfnamefont{M.}~\bibnamefont{Hayakawa}},
  \bibinfo{author}{\bibfnamefont{T.}~\bibnamefont{Kinoshita}},
  \bibnamefont{and} \bibinfo{author}{\bibfnamefont{M.}~\bibnamefont{Nio}},
  \bibinfo{journal}{Phys. Rev. Lett.} \textbf{\bibinfo{volume}{99}},
  \bibinfo{pages}{110406} (\bibinfo{year}{2007}).

\bibitem[{\citenamefont{Aoyama et~al.}(2012)\citenamefont{Aoyama, Hayakawa,
  Kinoshita, and Nio}}]{Aoyama12}
\bibinfo{author}{\bibfnamefont{T.}~\bibnamefont{Aoyama}},
  \bibinfo{author}{\bibfnamefont{M.}~\bibnamefont{Hayakawa}},
  \bibinfo{author}{\bibfnamefont{T.}~\bibnamefont{Kinoshita}},
  \bibnamefont{and} \bibinfo{author}{\bibfnamefont{M.}~\bibnamefont{Nio}},
  \bibinfo{journal}{Phys. Rev. Lett.} \textbf{\bibinfo{volume}{109}},
  \bibinfo{pages}{111807} (\bibinfo{year}{2012}).

\bibitem[{\citenamefont{Pachucki}(2008)}]{Pachucki08}
\bibinfo{author}{\bibfnamefont{K.}~\bibnamefont{Pachucki}},
  \bibinfo{journal}{Phys. Rev. A} \textbf{\bibinfo{volume}{78}},
  \bibinfo{pages}{012504} (\bibinfo{year}{2008}).

\bibitem[{\citenamefont{Nefiodov et~al.}(2002)\citenamefont{Nefiodov, Plunien,
  and Soff}}]{Nefiodov02}
\bibinfo{author}{\bibfnamefont{A.~V.} \bibnamefont{Nefiodov}},
  \bibinfo{author}{\bibfnamefont{G.}~\bibnamefont{Plunien}}, \bibnamefont{and}
  \bibinfo{author}{\bibfnamefont{G.}~\bibnamefont{Soff}},
  \bibinfo{journal}{Phys. Rev. Lett.} \textbf{\bibinfo{volume}{89}},
  \bibinfo{pages}{081802} (\bibinfo{year}{2002}).

\bibitem[{\citenamefont{Jentschura et~al.}(2006)\citenamefont{Jentschura,
  Czarnecki, Pachucki, and Yerokhin}}]{Jentschura06}
\bibinfo{author}{\bibfnamefont{U.~D.} \bibnamefont{Jentschura}},
  \bibinfo{author}{\bibfnamefont{A.}~\bibnamefont{Czarnecki}},
  \bibinfo{author}{\bibfnamefont{K.}~\bibnamefont{Pachucki}}, \bibnamefont{and}
  \bibinfo{author}{\bibfnamefont{V.~A.} \bibnamefont{Yerokhin}},
  \bibinfo{journal}{Int. J. Mass. Spectrom.} \textbf{\bibinfo{volume}{251}},
  \bibinfo{pages}{102 } (\bibinfo{year}{2006}).

\bibitem[{\citenamefont{Czarnecki et~al.}(1996)\citenamefont{Czarnecki, Krause,
  and Marciano}}]{Czarnecki96}
\bibinfo{author}{\bibfnamefont{A.}~\bibnamefont{Czarnecki}},
  \bibinfo{author}{\bibfnamefont{B.}~\bibnamefont{Krause}}, \bibnamefont{and}
  \bibinfo{author}{\bibfnamefont{W.~J.} \bibnamefont{Marciano}},
  \bibinfo{journal}{Phys. Rev. Lett.} \textbf{\bibinfo{volume}{76}},
  \bibinfo{pages}{3267} (\bibinfo{year}{1996}).

\bibitem[{\citenamefont{Nomura and Teubner}(2013)}]{Nomura13}
\bibinfo{author}{\bibfnamefont{D.}~\bibnamefont{Nomura}} \bibnamefont{and}
  \bibinfo{author}{\bibfnamefont{T.}~\bibnamefont{Teubner}},
  \bibinfo{journal}{Nucl. Phys. B} \textbf{\bibinfo{volume}{867}},
  \bibinfo{pages}{236} (\bibinfo{year}{2013}), ISSN \bibinfo{issn}{0550-3213}.

\bibitem[{\citenamefont{Kurz et~al.}(2014)\citenamefont{Kurz, Liu, Marquard,
  and Steinhauser}}]{Kurz14}
\bibinfo{author}{\bibfnamefont{A.}~\bibnamefont{Kurz}},
  \bibinfo{author}{\bibfnamefont{T.}~\bibnamefont{Liu}},
  \bibinfo{author}{\bibfnamefont{P.}~\bibnamefont{Marquard}}, \bibnamefont{and}
  \bibinfo{author}{\bibfnamefont{M.}~\bibnamefont{Steinhauser}},
  \bibinfo{journal}{Phys. Lett. B} \textbf{\bibinfo{volume}{734}},
  \bibinfo{pages}{144} (\bibinfo{year}{2014}).

\bibitem[{\citenamefont{Prades et~al.}(2010)\citenamefont{Prades, de~Rafael,
  and Vainshtein}}]{Prades10}
\bibinfo{author}{\bibfnamefont{J.}~\bibnamefont{Prades}},
  \bibinfo{author}{\bibfnamefont{E.}~\bibnamefont{de~Rafael}},
  \bibnamefont{and}
  \bibinfo{author}{\bibfnamefont{A.}~\bibnamefont{Vainshtein}}
  (\bibinfo{publisher}{World Scientific}, \bibinfo{address}{Singapore},
  \bibinfo{year}{2010}), vol.~\bibinfo{volume}{20} of
  \emph{\bibinfo{series}{Advanced Series on Directions in High Energy
  Physics}}, chap.~\bibinfo{chapter}{9}, pp. \bibinfo{pages}{303--317}.

\bibitem[{\citenamefont{Sick}(2015)}]{Sick15}
\bibinfo{author}{\bibfnamefont{I.}~\bibnamefont{Sick}},
  \bibinfo{journal}{Journal of Physical and Chemical Reference Data}
  \textbf{\bibinfo{volume}{44}}, \bibinfo{pages}{031213}
  (\bibinfo{year}{2015}).

\bibitem[{\citenamefont{Czarnecki et~al.}(2000)\citenamefont{Czarnecki,
  Melnikov, and Yelkhovsky}}]{Czarnecki00}
\bibinfo{author}{\bibfnamefont{A.}~\bibnamefont{Czarnecki}},
  \bibinfo{author}{\bibfnamefont{K.}~\bibnamefont{Melnikov}}, \bibnamefont{and}
  \bibinfo{author}{\bibfnamefont{A.}~\bibnamefont{Yelkhovsky}},
  \bibinfo{journal}{Phys. Rev. A} \textbf{\bibinfo{volume}{63}},
  \bibinfo{pages}{012509} (\bibinfo{year}{2000}).

\bibitem[{\citenamefont{Mohr and Taylor}(2005)}]{CODATA2005}
\bibinfo{author}{\bibfnamefont{P.~J.} \bibnamefont{Mohr}} \bibnamefont{and}
  \bibinfo{author}{\bibfnamefont{B.~N.} \bibnamefont{Taylor}},
  \bibinfo{journal}{Rev. Mod. Phys.} \textbf{\bibinfo{volume}{77}},
  \bibinfo{pages}{1} (\bibinfo{year}{2005}).

\bibitem[{\citenamefont{Martin and Zalubas}(1983)}]{Martin83}
\bibinfo{author}{\bibfnamefont{W.~C.} \bibnamefont{Martin}} \bibnamefont{and}
  \bibinfo{author}{\bibfnamefont{R.}~\bibnamefont{Zalubas}},
  \bibinfo{journal}{J. Phys. Chem. Ref. Data} \textbf{\bibinfo{volume}{12}},
  \bibinfo{pages}{323} (\bibinfo{year}{1983}).

\bibitem[{\citenamefont{Mohr et~al.}(2012)\citenamefont{Mohr, Taylor, and
  Newell}}]{CODATA2012}
\bibinfo{author}{\bibfnamefont{P.~J.} \bibnamefont{Mohr}},
  \bibinfo{author}{\bibfnamefont{B.~N.} \bibnamefont{Taylor}},
  \bibnamefont{and} \bibinfo{author}{\bibfnamefont{D.~B.}
  \bibnamefont{Newell}}, \bibinfo{journal}{Rev. Mod. Phys.}
  \textbf{\bibinfo{volume}{84}}, \bibinfo{pages}{1527} (\bibinfo{year}{2012}).

\bibitem[{\citenamefont{Sturm et~al.}(2017)\citenamefont{Sturm, Vogel,
  K\"ohler-Langes, Quint, Blaum, and Werth}}]{Sturm17}
\bibinfo{author}{\bibfnamefont{S.}~\bibnamefont{Sturm}},
  \bibinfo{author}{\bibfnamefont{M.}~\bibnamefont{Vogel}},
  \bibinfo{author}{\bibfnamefont{F.}~\bibnamefont{K\"ohler-Langes}},
  \bibinfo{author}{\bibfnamefont{W.}~\bibnamefont{Quint}},
  \bibinfo{author}{\bibfnamefont{K.}~\bibnamefont{Blaum}}, \bibnamefont{and}
  \bibinfo{author}{\bibfnamefont{G.}~\bibnamefont{Werth}},
  \bibinfo{journal}{Atoms} \textbf{\bibinfo{volume}{5}} (\bibinfo{year}{2017}).

\bibitem[{\citenamefont{Sturm et~al.}(2013{\natexlab{b}})\citenamefont{Sturm,
  Werth, and Blaum}}]{Sturm13AP}
\bibinfo{author}{\bibfnamefont{S.}~\bibnamefont{Sturm}},
  \bibinfo{author}{\bibfnamefont{G.}~\bibnamefont{Werth}}, \bibnamefont{and}
  \bibinfo{author}{\bibfnamefont{K.}~\bibnamefont{Blaum}},
  \bibinfo{journal}{Ann. Phys.} \textbf{\bibinfo{volume}{525}},
  \bibinfo{pages}{620} (\bibinfo{year}{2013}{\natexlab{b}}).

\bibitem[{\citenamefont{Yerokhin}(2017)}]{Yerokhin17}
\bibinfo{author}{\bibfnamefont{V.~A.} \bibnamefont{Yerokhin}},
  \bibinfo{howpublished}{private communication} (\bibinfo{year}{2017}).

\bibitem[{\citenamefont{Streubel et~al.}(2014)\citenamefont{Streubel, Eronen,
  H{\"o}cker, Ketter, Schuh, Van~Dyck, and Blaum}}]{Streubel14}
\bibinfo{author}{\bibfnamefont{S.}~\bibnamefont{Streubel}},
  \bibinfo{author}{\bibfnamefont{T.}~\bibnamefont{Eronen}},
  \bibinfo{author}{\bibfnamefont{M.}~\bibnamefont{H{\"o}cker}},
  \bibinfo{author}{\bibfnamefont{J.}~\bibnamefont{Ketter}},
  \bibinfo{author}{\bibfnamefont{M.}~\bibnamefont{Schuh}},
  \bibinfo{author}{\bibfnamefont{R.~S.} \bibnamefont{Van~Dyck}},
  \bibnamefont{and} \bibinfo{author}{\bibfnamefont{K.}~\bibnamefont{Blaum}},
  \bibinfo{journal}{Appl. Phys. B} \textbf{\bibinfo{volume}{114}},
  \bibinfo{pages}{137} (\bibinfo{year}{2014}).

\bibitem[{\citenamefont{Shabaev et~al.}(2006)\citenamefont{Shabaev, Glazov,
  Oreshkina, Volotka, Plunien, Kluge, and Quint}}]{Shabaev06}
\bibinfo{author}{\bibfnamefont{V.~M.} \bibnamefont{Shabaev}},
  \bibinfo{author}{\bibfnamefont{D.~A.} \bibnamefont{Glazov}},
  \bibinfo{author}{\bibfnamefont{N.~S.} \bibnamefont{Oreshkina}},
  \bibinfo{author}{\bibfnamefont{A.~V.} \bibnamefont{Volotka}},
  \bibinfo{author}{\bibfnamefont{G.}~\bibnamefont{Plunien}},
  \bibinfo{author}{\bibfnamefont{H.-J.} \bibnamefont{Kluge}}, \bibnamefont{and}
  \bibinfo{author}{\bibfnamefont{W.}~\bibnamefont{Quint}},
  \bibinfo{journal}{Phys. Rev. Lett.} \textbf{\bibinfo{volume}{96}},
  \bibinfo{pages}{253002} (\bibinfo{year}{2006}).

\bibitem[{\citenamefont{Yerokhin
  et~al.}(2016{\natexlab{a}})\citenamefont{Yerokhin, Berseneva, Harman,
  Tupitsyn, and Keitel}}]{Yerokhin16}
\bibinfo{author}{\bibfnamefont{V.~A.} \bibnamefont{Yerokhin}},
  \bibinfo{author}{\bibfnamefont{E.}~\bibnamefont{Berseneva}},
  \bibinfo{author}{\bibfnamefont{Z.}~\bibnamefont{Harman}},
  \bibinfo{author}{\bibfnamefont{I.~I.} \bibnamefont{Tupitsyn}},
  \bibnamefont{and} \bibinfo{author}{\bibfnamefont{C.~H.}
  \bibnamefont{Keitel}}, \bibinfo{journal}{Phys. Rev. Lett.}
  \textbf{\bibinfo{volume}{116}}, \bibinfo{pages}{100801}
  (\bibinfo{year}{2016}{\natexlab{a}}).

\bibitem[{\citenamefont{Yerokhin
  et~al.}(2016{\natexlab{b}})\citenamefont{Yerokhin, Berseneva, Harman,
  Tupitsyn, and Keitel}}]{Yerokhin16PRA}
\bibinfo{author}{\bibfnamefont{V.~A.} \bibnamefont{Yerokhin}},
  \bibinfo{author}{\bibfnamefont{E.}~\bibnamefont{Berseneva}},
  \bibinfo{author}{\bibfnamefont{Z.}~\bibnamefont{Harman}},
  \bibinfo{author}{\bibfnamefont{I.~I.} \bibnamefont{Tupitsyn}},
  \bibnamefont{and} \bibinfo{author}{\bibfnamefont{C.~H.}
  \bibnamefont{Keitel}}, \bibinfo{journal}{Phys. Rev. A}
  \textbf{\bibinfo{volume}{94}}, \bibinfo{pages}{022502}
  (\bibinfo{year}{2016}{\natexlab{b}}).

\end{thebibliography}
\end{document}